\documentstyle[preprint,prd,aps,12pt]{revtex}
\preprint{
$
\begin{array}{r}
\text{LAVAL-PHY-94-13} \\
\text{McGILL/94-26} \\
\text{SPhT 94/072} \\
\end{array}
$
}
\tighten

\begin{document}
\author{M. de Montigny}
\address{Physics Department, McGill University \\
Montreal, Canada, H3A 2T8}
\author{L. Marleau}
\address{D\'epartement de Physique, Universit\'e Laval\\
Qu\'ebec,Canada,G1K 7P4\\
and\\
Service de\ Physique\ Th\'eorique,CEA-Saclay\\
F-91191,Gif-sur-Yvette,France}
\author{G. Simon}
\address{D\'epartement de Physique, Universit\'e Laval\\
Qu\'ebec, Canada, G1K 7P4}
\title{Production of Pairs of Sleptoquarks in Hadron Colliders}
\date{June 1994}
\maketitle
\draft

\begin{abstract}
We calculate the cross section for the production of pairs of scalar
leptoquarks (sleptoquarks) in a supersymmetric $E_6$ model, at hadron
colliders. We estimate higher order corrections by including $\pi^2$ terms
induced by soft-gluon corrections. Discovery bounds on the sleptoquark mass
are estimated at collider energies of 1.8, 2, and 4 TeV (Tevatron), and 16
TeV (LHC).
\end{abstract}

\pacs{PACS numbers: 14.80.-j, 11.25.Mj. }

A natural explanation for the proliferation of fermions, and their pattern
of masses and mixing angles, is to assume that the current elementary
particles are composite. In various ``preonic'' models \cite{af}, quarks and
leptons have some common constituents. Among the particles contained in
these models, {\it leptoquarks} are exotic particles which have both a
nonzero leptonic and baryonic number. They also appear in many extensions of
the Standard Model \cite{gutlq}. They can decay directly into a quark-lepton
pair, which is a new feature, since there are no quark-lepton-boson
interactions in the Standard Model. Here we are concerned with the
leptoquarks that come from a supersymmetric grand unified $E_6$ theory (the
low energy limit of an $E_8\otimes E_8$ heterotic string theory \cite{string}%
), where they --- together with their supersymmetric scalar partners,
sleptoquarks --- are fractionally charged color triplets. For constraints on
the parameters of these superstring-inspired models, see, for example, Ref.
\cite{ceegn} or for a more extensive review, Ref. \cite{hr}.

In principle, the best place to look for a leptoquark signal would be at $ep$
colliders \cite{ep} since they can be produced directly via the
lepton-quark-sleptoquark coupling (called from now on the {\it Yukawa}
coupling). However, for small couplings (or large leptoquark masses), hadron
colliders seem more appropriate since leptoquarks can still be pair produced
through their strong interactions for arbitrarily low Yukawa couplings.
Experiments at LEP have already imposed a lower bound of $45$ GeV on the
leptoquark mass \cite{lep}. Results from HERA have ruled out masses below $%
180$ GeV \cite{hera} for a Yukawa coupling of electromagnetic strength,
whereas searches at $p\overline{p}$ colliders have set these bounds to $113$
GeV (for a branching ratio BR$=1$), or $80$ GeV (BR$=0.5$) \cite{abe}. More
recently, experiments both at HERA and the Tevatron have strengthened these
bounds reaching with a $95\%$ confidence level ,  $240$ GeV for HERA \cite
{hapke} and respectively  $133$ GeV and $120$ GeV at the Tevatron  \cite{d0}%
. Note that HERA bounds depends on the value of the Yukawa coupling whereas
the hadron collider results are somewhat insensitive to this parameter.

Low energy data ({\it e.g.} atomic parity violation, \dots) also impose very
strict bounds on leptoquark masses. Leurer has updated a previous analysis
by Buchm\"uller and Wyler, and obtained bounds that restrict the leptoquarks
(with Yukawa coupling equal to the electromagnetic strength) to have masses
larger than $600$ GeV or $630$ GeV for leptoquarks that couple to RH quarks,
and above $1040$ GeV, $440$ GeV, and $750$ GeV for the $SU(2)_W$ scalar,
doublet and triplet leptoquarks, respectively, that couple to LH quarks \cite
{bounds}. These are ``unavoidable'' bounds, in the sense that they are
independent of the following three assumptions, which are used to circumvent
other constraints: (1) leptoquarks do not also couple to diquarks, (2)
leptoquarks couple chirally ({\it i.e.} to one quark chirality at a time),
and (3) leptoquarks couple diagonally ({\it i.e.} to a single fermion
generation at a time), but they remain heavily dependent on the strength of
the Yukawa coupling. Finally, some model-independent bounds are also
discussed by Davidson et al. in Ref. \cite{bounds}. Hence, it is quite
possible that leptoquarks could escape detection at $ep$ colliders but could
still be detectable at hadron colliders. Also, in practice, the energy
available in hadron colliders is greater than that in $ep$ colliders, which
would in principle extend the search to larger values of the leptoquark mass.

The purpose of this brief report is to reanalyze the production of pairs of
sleptoquarks in hadron colliders \cite{hewpak,dm90} in view of the most
recent data and experimental situation. In previous work, we calculated the
total cross section at center-of-mass energies of $2$, $16$, and $40$ TeV
(corresponding to the Fermilab Tevatron, the CERN LHC, and the late SSC,
respectively), for sleptoquark masses up to $400$ GeV. We extend here our
calculation of the cross section to sleptoquark masses of up to $2600$ GeV,
and we add the results for the (current and newly proposed) Tevatron
energies $\sqrt{s}=1.8$ TeV and $4$ TeV. (We keep the SSC energies in order
to compare with our previous results.) The soft-gluon $\pi^2$ term
corrections are included. We also have taken special care to reduce
uncertainties in the numerical calculations by performing a more reliable
numerical integration (using VEGAS \cite{vegas}), and by taking a more
recent set of distribution functions.

Let us briefly review the model and our calculations. (More details are
given in Refs. \cite{dm90}, and references therein). In the  supersymmetric $%
E_6$ model, the leptoquark $D$ is an exotic colored particle which lies in
the {\bf 27} fermionic multiplet, and the sleptoquark that we consider is
its scalar superpartner. If we restrict our study to the first generation of
fermions, then the Yukawa interactions take the form:
\begin{equation}
\label{L}{\cal L}_Y=\lambda_L{\tilde D}^{c*}\left( e_Lu_L+\nu_Ld_L\right)
+\lambda_R{\tilde D}e_L^cu_L^c+\text{h.c.}
\end{equation}
where $c$ denotes the charge conjugate state, and ${\tilde D}$ is the scalar
superpartner of $D$. We assume the $\lambda$'s to be independent and
arbitrary. We analyze the pair production of sleptoquarks which arise from
two subprocesses: (1) quark-antiquark annihilation $(u_R+u_L^c\rightarrow {%
\tilde D}+{\tilde D}^{*}$ and $u_L+u_R^c\rightarrow {\tilde D}^{c*}+{\tilde D%
}^c)$, and (2) gluon fusion $(g+g\rightarrow {\tilde D}+{\tilde D}^{*}$ and $%
g+g\rightarrow {\tilde D}^{c*}+{\tilde D}^c)$. The subprocess (1) occurs in $%
s$-channel (through the exchange of a virtual gluon) and in $t$-channel
(virtual electron). The subprocess (2) arises via color gauge interactions
from the trilinear term $gDD$ in the $s$-channel (through the exchange of a
gluon) and in the $t$- and $u$-channels (exchange of virtual sleptoquarks),
and from the quartic term $ggDD$ in which two gluons annihilate to produce
directly a pair of sleptoquarks. Note that, to lowest order, the gluon
fusion process does not depend on the Yukawa interactions. The details of
the calculation of the amplitudes and the cross sections are given in Refs.
\cite{dm90}.

We also estimate $\pi^2$ terms, which are the soft-gluon corrections that
arise from the regularization of either collinear or infrared singularities,
when a timelike momentum transfer is involved in the process \cite{dm89}.
The first-order corrections to a subprocess involving massless particles
contain an infrared singularity of the form
\begin{equation}
\label{ir}{\text{Re}}\left[ {\frac 1{\epsilon^2}}\left( -{\frac{q^2}{\mu^2} }%
\right)^{-\epsilon }\right] ={\frac 1{\epsilon^2}}-{\frac 1\epsilon }{\
\text{Re}}\left[ {\text{ln}}\left( -{\frac{q^2}{\mu^2}}\right) \right] +{\
\frac 12}{\text{Re}}\left[ {\text{ln}}^2\left( -{\frac{q^2}{\mu^2}}\right)
\right] +\cdots
\end{equation}
where $\mu^2$ is the renormalization point. The $\pi^2$ term is generated by
the last term,
\begin{equation}
\label{irps}{\text{Re}}\left[ {\text{ln}}^2\left( -{\frac{q^2}{\mu^2}}
\right) \right] ={\text{ln}}^2\left( {\frac{q^2}{\mu^2}}\right) -\pi^2.
\end{equation}
Note that $\pi^2$ terms arise only when $q^2$ is timelike. Unless there is a
suppression of the $\pi^2$ terms, their contribution is large and cannot be
neglected. It is usually expressed by means of the so-called $K$-{\it factor}%
. In general, we expect $\pi^2$ terms to appear to all orders in $\alpha_S$.
Low-order results in many QCD processes indicate that the summation of the
ensuing large corrections into an exponential could take place, as is the
case for electromagnetic interactions. It is important to remark that not
all of the first-order correction diagrams lead to a $\pi^2$ term. (The
detailed investigation of the soft-gluon corrections leading to a $K$-factor
through $\pi^2$ term is described in Ref. \cite{dm90}, for $q\overline{q}$
and $gg$ subprocesses. The analogous study for other leptoquark processes is
given in Ref. \cite{dm89}.)

We now display the contributions to the total cross section for each of the
two subprocesses mentioned previously, including the $K$-factors coming from
soft-gluon corrections. For the Born term only, one just has to set $K=1$.
The contribution from the $q{\overline{q}}$ annihilation is:
\begin{equation}
\label{qq}\sigma_{AB}=\int_{x_{min}/x}^1dy\int_{x_{min}}^1dxK_{AB}
\alpha_A\alpha_B{\tilde\sigma}_{AB}\left[ G_{u/p}(x,Q^2)G_{\overline{u}/ {%
\overline{p}}}(y,Q^2)+(1\leftrightarrow 2,p\leftrightarrow {\overline{p}}
)\right] ,
\end{equation}
with $x_{min}=4(M_D^2+p_T^2)/s$ (we take $p_T=10$ GeV), and $Q^2={\hat s}/2$
(${\hat s}=xys$). Here $AB$ stands for $SS$, $YY$, or $SY$ ({\it i.e.}
purely strong, purely Yukawa, and mixed subprocesses, respectively) where $%
\alpha_S$ and $\alpha_Y$ are the QCD and Yukawa couplings, respectively. We
set $\alpha_Y=\alpha_{em}$ in the numerical calculations. Finally, $%
G_{a/h}(x,Q^2)$ stands for the distribution function associated to the
parton $a$ in the hadron $h$, with scaling variable $x$ and momentum scale $%
Q^2$. We use the parametrization set B2 of Morfin and Tung \cite{mt}, which
is based on more recent nucleon data than the parametrization we used in our
previous work. Defining $\chi =M_D^2/{\hat s}$ and $\eta =\sqrt{1-4M_D^2/{%
\hat s}}$, the purely strong, mixed, and purely Yukawa contributions are,
respectively,
\begin{equation}
\label{ss}{\tilde \sigma }_{SS}={\frac{2\pi }{27{\hat s}}}\eta^3,
\end{equation}
\begin{equation}
\label{sy}{\tilde \sigma }_{SY}={\frac{2\pi }{36{\hat s}}}\left[ \eta
(1-2\chi )+2\chi^2{\text{ln}}\left( {\frac{1-\eta -2\chi }{1+\eta -2\chi }}
\right) \right] ,
\end{equation}

\begin{equation}
\label{yy}{\tilde\sigma}_{YY}= {\frac \pi{8{\hat s}}} \left[ -2\eta +(2\chi
-1) {\text{ln}} \left( {\frac{1-\eta -2\chi} {1+\eta -2\chi}}\right) \right]
{}.
\end{equation}

The $K$-factors are found to be

\begin{equation}
\label{Kqq}K_{SS}=K_{SY}=1+\pi \alpha_S\left( C_F-{\frac 12}C_A\right) =1-{\
\frac 16}\pi \alpha_S,\text{ \quad and\quad }K_{YY}=1.
\end{equation}
Here $C_F={\frac 43}$ and $C_A=3$ are the Casimir operators for the $SU(3)$
fundamental and adjoint representations, respectively. (Later on, $K_{SS}$
and $K_{SY}$ will be denoted generically as $K_{qq}$.)

For the gluon fusion subprocess, the integrated cross section is
\begin{equation}
\sigma_{gg}=\int_{x_{min}/x}^1dy\int_{x_{min}}^1dxK_{gg}\alpha_S^2{\tilde
\sigma }_{gg}G_{g/p}(x,Q^2)G_{g/{\overline p}}(y,Q^2),
\end{equation}
with
\begin{equation}
\label{intgg}{\tilde \sigma }_{gg}={\frac \pi {6{\hat s}}}\left[ \left( {\
\frac 58}+{\frac{31}4}\chi \right) \eta +\left( 4+\chi \right) \chi {\text{%
ln }}\left( {\frac{1-\eta }{1+\eta }}\right) \right] .
\end{equation}
The values for $x_{min},\ p_T,\ Q^2$ are the same as in (\ref{qq}). Here,
the $K$-factor induces a much larger correction to the Born term,
\begin{equation}
\label{Kgg}K_{gg}=1+{\frac 12}\pi\alpha_SC_F +\cdots =1+{\frac 23}\pi
\alpha_S+\cdots
\end{equation}
The cross sections (\ref{ss}-\ref{yy}) and (\ref{intgg}) are similar to
those in \cite{hls} (for $q{\overline q}\rightarrow {\tilde q} {\tilde q}$, $%
gg\rightarrow {\tilde q}{\tilde q}$), except for a factor. For the Born
term, one just replaces $K$ by $1$ in (\ref{Kqq}) and (\ref{Kgg}). Fig. \ref
{figone} displays the total cross section, that is, the sum of $gg$ fusion
and $q{\overline q}$ contributions, with and without corrections.

Higher order effects (only the $\pi^2$ terms are included here) are easy to
estimate. {}From (\ref{Kqq}) the $\pi^2$ corrections for the quark-antiquark
annihilation are small and negative ({\it i.e.} they slightly suppress the
cross section). For sleptoquark masses in the range $100$ GeV $<M_D<2600$
GeV, the strong coupling constant $\alpha_S$ lies between $0.13$ and $0.09$,
which corresponds to $K_{qq}=0.932$ and $0.953$, respectively. Here the
cross section is suppressed by $7$ to $4\%$, depending on the relative
importance of the different channels. For the gluon fusion process (\ref{Kgg}%
), the cross section is significantly increased with $K_{gg}=1.272$ and $%
1.189$ for $\alpha_S=0.13$ and $0.09$, respectively. From Fig. \ref{figone},
we see that the $\pi^2$ term leads to a slight increase of the cross section
for the small values of the leptoquark mass, and that the signal  is
suppressed for large values. This is due to the fact that, at high energies,
the gluon fusion subprocess--which  undergoes bigger corrections--
dominates in the region of low leptoquark mass. The value of the mass for
which the  two curves (Born and corrected) intersect increases with  the
center-of-mass energy.  These corrections may be crucial if the values of $%
\alpha_Y$ and $M_D$ are such that we are close to the detection threshold.
It should be noted that the $K$-factors found here are only part of the
first order corrections. However, whereas one expects $K_{qq}$ to be of the
same order of magnitude as the rest of the first order corrections, it
appears that $K_{gg}$ is the dominant part of the gluon fusion corrections.

We expect the final leptoquarks to decay into a quark and a lepton. The most
interesting signature for ${\tilde D}{\tilde D}^{*}$ pair is the production
of 2 jets $+ l^{+}l^{-}$. The standard background would come from $p%
\overline{p}\rightarrow Q\overline{Q}$, where $Q$ is a heavy quark which
decays semileptonically into 2 jets $+ l^{+}l^{-}+\nu \overline{\nu}$, and
will in general involve some missing $p_T$. It is easy to see that the
subprocesses involved in $Q\overline{Q}$ production get similar $\pi^2$
corrections (i.e. a factor $K_{qq}$ for $q\overline{q}\rightarrow Q\overline{%
Q}$ and a factor $K_{gg}$ for $gg\rightarrow Q\overline{Q}$). This could
slightly modify the  signal-to-background ratio in a manner which depends on
parameters such as the masses of sleptoquarks and heavy quarks and the
relative importance of each subprocess. For example, if gluon fusion is the
dominant process in both sleptoquark and heavy quark pair production, the
signal-to-background ratio will not be affected by the soft-gluon
corrections, but the total cross section will be appreciably enhanced.

Finally, we discuss the detection possibilities, taking into account the
expected luminosities at the various colliders considered in this work. In
all cases, we assume that the discovery limit is $20$ events, and we base
our estimates on the corrected cross sections ({\it i.e.} including $K$%
-factors). We have obtained the results of Fig. \ref{figone} by putting $%
\alpha_Y=\alpha_{em}$. For the sake of comparison we give also an estimate
for the case of a very small Yukawa coupling as well, by setting $\alpha_Y=0$%
. Experiments at Fermilab expect to gather about $100$ pb$^{-1}$ during the
current Tevatron run (for which $\sqrt{s}=1.8$ TeV), which should last until
the middle of 1995. With the discovery limit mentioned above, the mass reach
would be $184$ GeV (if $\alpha_Y=0$ rather than $\alpha_{em}$, then it is $%
181$ GeV). If we assume a luminosity equal to $500$ pb$^{-1}$ at $\sqrt{s}=2$
TeV, then this number would increase by an appreciable amount: up to $251$
GeV ($247$ GeV for $\alpha_Y=0$). The design luminosity at LHC is expected
to be $10^{34}$ cm$^{-2}$ s$^{-1}$. A run of one year would then give a
luminosity of $100$ fb$^{-1}$, providing a search limit of $2110$ GeV (or $%
2070$ GeV for $\alpha_Y=0$). Fermilab has recently started discussing an
upgrade of the center-of-mass energy to $4$ TeV, and the numbers used by
different people as expected luminosities are $1$ fb$^{-1}$ and $10$ fb$^{-1}
$. The discovery of $20$ events would be reached for masses not exceeding $%
434$ ($428$) GeV and $597$ ($590$) GeV, respectively (numbers in parentheses
are for $\alpha_Y=0$). At the SSC, with a hypothetical luminosity of $10$ fb$%
^{-1}$, the same discovery limit would lead to a mass reach of $2565$ GeV
for $\alpha_Y=\alpha_{em}$, and $2525$ GeV for $\alpha_Y=0$. Actually,
changing from $\alpha_Y=\alpha_{em}$ to $\alpha_Y=0$ has an overall effect
of only a few percent on the discovery limits. This means that the strong
processes dominate even at high energies where the ratio $%
\alpha_{em}(Q^2)/\alpha_S(Q^2)$ is not that small.

\bigskip
We are indebted to N. Hadley for information on the experiments being
carried out at the Fermilab Tevatron, P. Labelle for discussions about our
numerical calculations, and D. London for reading the manuscript. L.M. is
grateful for the hospitality of the Service de Physique Th\'eorique
(CEA-Saclay), where part of this work was done. This research was supported
by  Fondation A.F.D.U. Qu\'ebec
(G.S.),  the Natural Sciences and Engineering Research Council of Canada,
and by the Fonds pour la Formation de Chercheurs et l'Aide \`a la Recherche
du Qu\'ebec.

\begin{figure}
\caption{
The total cross section for pair production of scalar leptoquarks at
$p{\overline p}$ colliders with a center-of-mass energy of $1.8, 2, 4, 16$
and $40$ TeV versus the mass of the leptoquark. The Yukawa coupling is
taken to be $\alpha_Y = \alpha_{em}$. Solid (broken) lines represent the
Born ($\pi^2$-corrected) cross sections. The bottom-left axis are valid
 for $1.8$ and $2$ TeV, and the top-right axis, for $4, 16$ and $40$
TeV.}

\label{figone}
\end{figure}

\end{document}